Nathan Chi[1]; Peter Washington[2], BA, MSc; Aaron Kline[1], BS; Arman Husic[1], BSc; Cathy Hou[3]; Chloe He[4], BS; Kaitlyn Dunlap[1], BA; Dennis Wall[1,4,5], PhD

[1] Department of Pediatrics (Systems Medicine), Stanford University, Stanford, CA, United States
[2] Department of Bioengineering, Stanford University, Stanford, CA, United States
[3] Department of Computer Science, Stanford University, Stanford, CA, United States
[4] Department of Biomedical Data Science, Stanford University, Stanford, CA, United States
[5] Department of Psychiatry and Behavioral Sciences, Stanford University, Stanford CA, United States

**Corresponding Author:**
Dennis Paul Wall, PhD
Division of Systems
Medicine Department of Biomedical Data Science
Stanford University
3145 Porter Dr
Palo Alto, CA, United States
Phone: 1 650 497 9193
Email: dpwall@stanford.edu

# Classifying Autism from Crowdsourced Semi-Structured Speech Recordings: A Machine Learning Approach

## Abstract

**Background:** Autism spectrum disorder (ASD) is a neurodevelopmental disorder which results in altered behavior, social development, and communication patterns. In past years, autism prevalence has tripled, with 1 in 54 children now affected. Given that traditional diagnosis is a lengthy, labor-intensive process which requires the work of trained physicians, significant attention has been given to developing systems that automatically diagnose and screen for autism.

**Objective:** Prosody abnormalities are among the most clear signs of autism, with affected children displaying speech idiosyncrasies (including echolalia, monotonous intonation, atypical pitch, and irregular linguistic stress patterns). In this work, we present a suite of machine learning approaches to detect autism in self-recorded speech audio captured from autistic and neurotypical (NT) children in home environments.

**Methods:** We consider three methods to detect autism in child speech: first, Random Forests trained on extracted audio features (including Mel-frequency cepstral coefficients); second, convolutional neural networks (CNNs) trained on spectrograms; and third, fine-tuned wav2vec 2.0—a state-of-the-art Transformer-based speech recognition model. We train our classifiers on our novel dataset of cellphone-recorded child speech audio curated from Stanford's *Guess What?* mobile game, an app designed to crowdsource videos of autistic and neurotypical children in a natural home environment.

**Results:** The Random Forest classifier achieves 70% accuracy, the fine-tuned wav2vec 2.0 model achieves 77% accuracy, and the CNN achieves 79% accuracy when classifying children's audio as either ASD or NT. We use five-fold cross-validation to evaluate model performance.

**Conclusions:** Our models were able to predict autism status when training on a varied selection of home audio clips with inconsistent recording qualities, which may be more generalizable to real world conditions. The results demonstrate that machine learning methods offer promise in detecting autism automatically from speech without specialized equipment.

## Introduction

Autism encompasses a spectrum of disorders characterized by delayed linguistic development, social interaction deficits, and behavioral impairments [1]. Autism prevalence has rapidly increased in recent years: according to the Centers for Disease Control and Prevention, autism rates have nearly tripled since 2000 to 1 in 54 children in 2016 [2]. In the United States alone, over 5 million individuals are affected [3], and nearly 75 million are affected worldwide. Despite the increasing prevalence of autism, access to diagnosis resources continues to be limited, with 83.86% of all American counties not having any [4]. These nationwide inadequacies in autism resources are compounded by the lengthy nature of diagnosis. On average, the delay from the time of first consultations with healthcare providers to the time of diagnosis is over 2 years. Such extensive delays often cause diagnosis at a later age (usually ≥ 4 years old) [5], which may result in greater lifelong impacts, including a higher likelihood of psychotropic medication use, lower IQ scores and reduced language aptitude [6, 65]. Given that timely autism identification and intervention has been shown to improve treatment success and social capabilities, research has focused on its early detection [7, 8, 31, 64, 65].

Although symptoms range from individual to individual, prosody abnormalities are among the most notable signs of autism spectrum disorder, with multiple studies suggesting that affected children display peculiarities including echolalia, monotonous intonation, and atypical pitch and linguistic stress patterns [9, 10, 11]. Given this, an effective AI sound classifier trained to detect speech abnormalities common in children with autism would be a valuable tool for autism diagnosis.

Prior research [12, 13] investigated prosodic disorders in children with autism, to varying degrees of success. Cho et. al [14] developed models that achieved 76% accuracy on a dataset of recorded interviews between children and unfamiliar adults, trained on data recorded at a consistent location using a specialized Biosensor device with four directional microphones. Similarly, Li et. al [15] achieve high accuracies when training on speech data recorded with multiple wireless microphones, providing high purity recordings at a central recording location (a hospital). However, both used data collected in centralized, unfamiliar locations with high-quality recording equipment. Such research, while promising, does not significantly accelerate the process of screening for autism, because they still require the use of specialized equipment and centralized recording locations to provide consistent audio quality, posing significant barriers to the widespread availability of automatic diagnosis tools. Additionally, interacting with strange adults in foreign environments could be stressful and possibly affect the behavior of children with autism, thus leading to observations that are not generalizable to the real world.

In this work, we propose a machine learning-based approach to predict signs of autism directly from self-recorded semi-structured home audio clips recording a child's natural behavior. We utilize Random Forests, CNNs, and fine-tuned wav2vec 2.0 models to identify differences in speech between autistic and neurotypical children. One strength of our approach is that our models are trained on mobile device audio recordings of varying audio quality. Therefore, unlike other studies, our approach does not necessitate specialized high-fidelity recording equipment. Additionally, we attempt to capture naturalistic speech patterns by recording children playing educational games with their parents in a low-stress home environment. Finally, our approach does not require a trained clinician to partake in conversation with the child. To our knowledge, our method is the first to screen for autism in a home environment without the use of specialized audio-recording devices.

## Methods

**Figure 1.** Overview of audio-based AI detection pipeline. First, the educational video game *Guess What?* crowdsources the recording of videos of neurotypical children and children with autism from consenting participants. Audio of children's speech is manually spliced from the videos; three models are trained on this audio data. The first is a Random Forest classifier, which utilizes an ensemble of independently trained decision trees. The second is a convolutional neural network (CNN). The third is a fine-tuned wav2vec 2.0 model. Model 1 takes commonly used speech recognition features as input, model 2 learns from spectrograms of the audio, and model 3 takes the raw audio data itself as input.

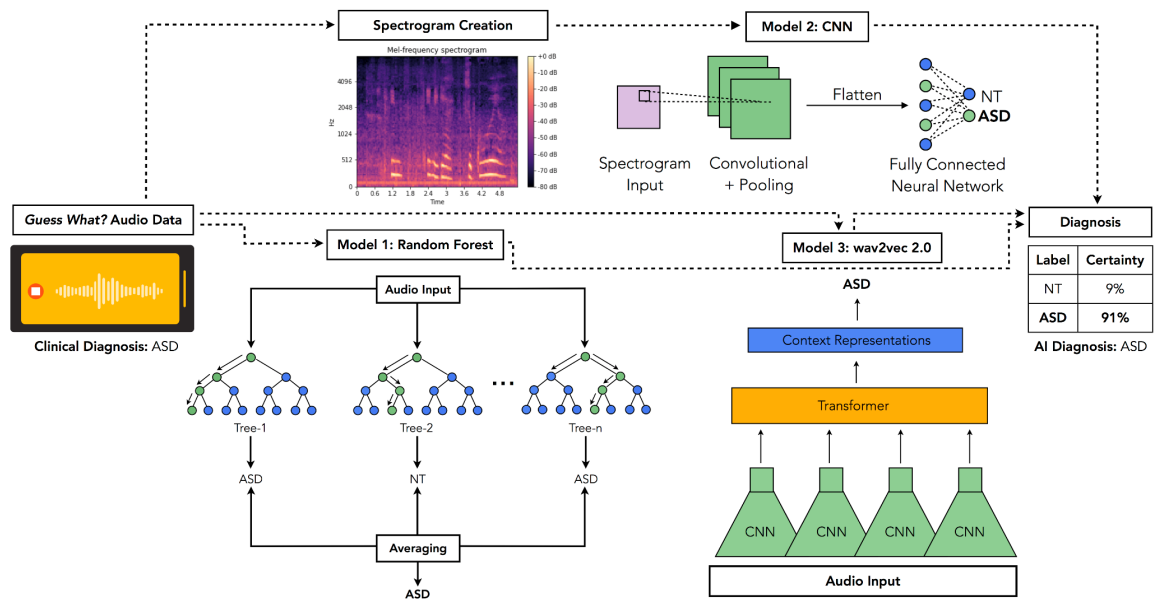

## Data Acquisition

### Process

We obtain audio data of neurotypical children and children with autism in a home environment through *Guess What?*, a mobile game designed for prosocial play and interaction at home between 2-8 year old developing children and their parents [16-20] (**Figure 1,** "*Guess What?* Audio Data"). During a game session, parents and children hoose either a charades game (acting out emotions, characters, sports, chores, or objects) or a simple quiz game (identifying colors, shapes, numbers, and word spellings). Children are directed to follow the rules of gameplay, while parents serve as game mediators. Throughout the session, parents record their children by placing their smartphones on their foreheads, with the front-facing camera oriented toward the child. After each 90-second session, parents are given the option to view their child's game session video recording and share it with our research team.

*Distribution*

We collected a total of 77 videos of 58 children participating in gameplay, recorded in the span of four years from 2018 to 2021. The participants range in age from 3 to 12 and include 20 ASD children (19 male and 1 female) and 38 neurotypical (NT) children (15 male, 22 female, and 1 unspecified). Parents involved in the study consented to sharing their videos with our research team and provided their child's age, sex, and diagnosis.

*Advantages*

This pipeline offers several benefits over traditional diagnostic workflows. Since only a smartphone is necessary, in theory more children can be screened for autism at lower cost. Through *Guess What?*, a traditionally time-intensive healthcare process for diagnosis could potentially be shortened into a quick—and even enjoyable—process. Furthermore, children recorded at home may be more likely to behave in a naturalistic manner.

## Data Preprocessing

Home videos are naturally variable in quality; their data contains a number of irregularities that must be addressed prior to analysis. In particular, parents or children would often join in gameplay simultaneously, resulting in a variety of voices, sometimes overlapping with one another. This overlap of voices can complicate the isolation and extraction of the child's voice. In order to remove adult speech, we manually sampled only child speech from each video, ensuring that each resulting clip did not include any other voice other than the child's. Each child contributed a mean of 1.32 videos and 14.7 clips, resulting in a total dataset size of 850 audio clips, representing 425 ASD and 425 NT clips. The 850 clips were split into five folds, as shown in **Table 1**, in preparation for five-fold cross-validation. When creating the folds, we included the restriction that all clips spliced from a given child's video had to be included in the same fold, in order to prevent models from learning from child-specific recording idiosyncrasies, including environmental background noise and audio quality.

**Table 1.** Distribution of 850 audio clips across 5 folds. Each of the three models were trained on the same distribution of clips with five-fold cross-validation.

|     | Fold 0 | Fold 1 | Fold 2 | Fold 3 | Fold 4 |
| --- | --- | --- | --- | --- | --- |
| NT  | 87 | 87 | 81 | 83 | 87 |
| ASD | 87 | 87 | 81 | 83 | 87 |

## Classifiers

We investigated three different methods to predict autism from audio, each represented in **Figure 1**.

- **Random Forest:** We trained random forests on a set of audio features (MFCC coefficients, chroma features, root-mean-square, spectral centroids, spectral bandwidths, spectral rolloff, and zero crossing rates) typically used in traditional signal processing speech recognition. We also tried training other models (including logistic regression, Gaussian Naive Bayes, and AdaBoosting models), which did not perform as well. We implemented the Random Forest model in scikit-learn and used the following manually chosen hyperparameters: [$\max_{\text{depth}}$ = 20000, $n_{\text{estimators}}$ = 56, $\max_{\text{features}}$ = 15, $\min_{\text{samples split}}$ = 10, $\min_{\text{samples leaf}}$ = 20, $\min_{\text{weight fraction leaf}}$ = 0.1].
- **CNN:** We trained a CNN using spectrograms of our data as input [21, 22]. Our spectrograms were synthesized via the python package Librosa. **Figure 2** shows an example of the spectrograms used to train the CNN. The CNN, represented in **Figure 3**, consists of 9 layers each with alternating convolution and max pooling layers, as well as three Dense layers with L2 regularization penalty = 0.01. We investigated both training a small CNN (~8M parameters) from scratch and fine-tuning the image recognition model Inception v3 (with ~33M parameters) trained on ImageNet [23]. However, our 8M parameter CNN ultimately performed slightly better than the transfer learning approach, likely due to the irrelevance of ImageNet features to spectrograms. Our final CNN model, which we train for 15 epochs (until training performance stopped improving), has 8,724,594 parameters.
- **wav2vec 2.0**: We fine-tuned wav2vec 2.0, a state-of-the-art Transformer model pretrained on a self-supervised audio denoising task [24]. Although wav2vec 2 is typically used for speech-to-text decoding, prior work [25] has demonstrated its utility for suprasegmental tasks such as emotion prediction. We use the **facebook/wav2vec2-base** variant and fine-tune for 264 steps. The final model has 95 million parameters.

**Figure 2**. Mel-frequency spectrogram for a neurotypical child speech segment, spliced from a *Guess What?* gameplay video. This spectrogram was one of 850 used to train the 8M CNN model, which yielded the highest accuracy of the three best-performing models.

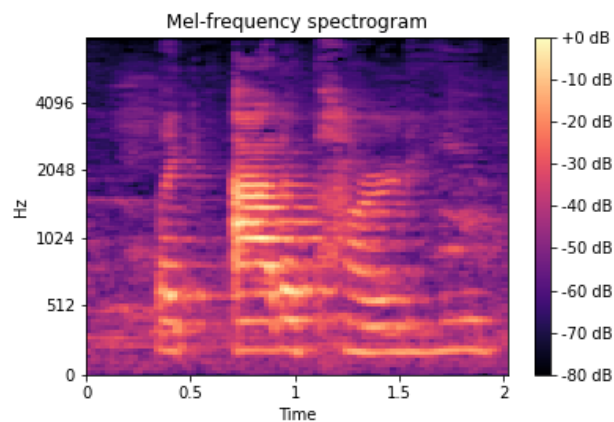

**Figure 3**. A) and B) represent the same ~8 million-parameter CNN model architecture. This architecture performed best out of all of our tested architectures, including a fine-tuned Inception v3 model. B) was in part created with the python package Visualkeras.

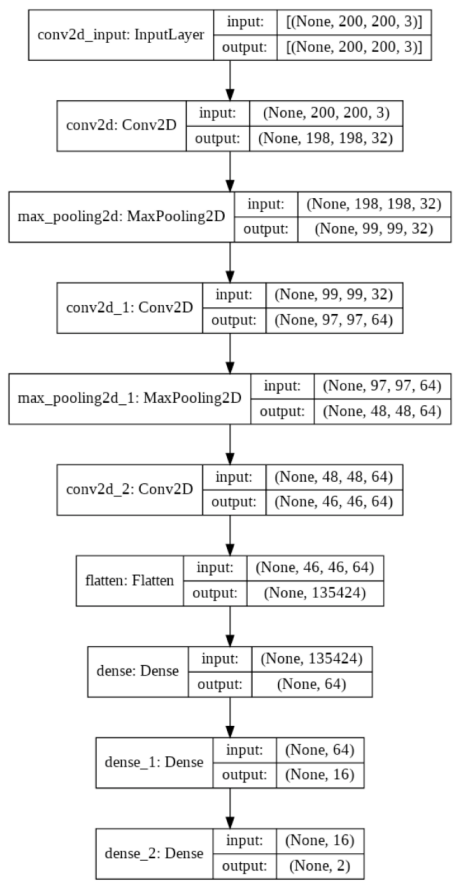
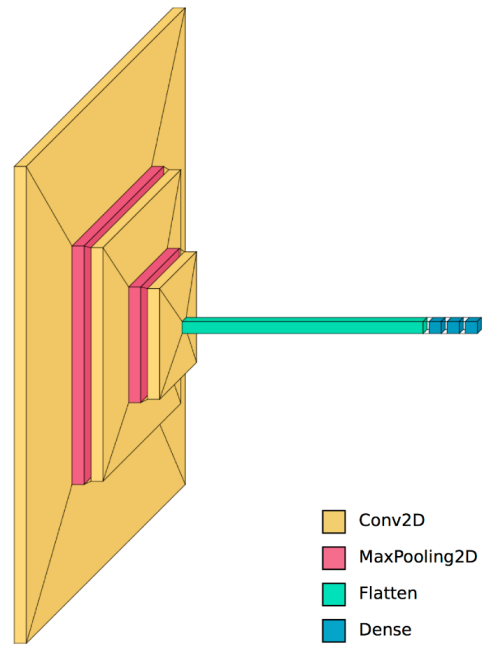

A) **8M CNN** Architecture

B) **8M CNN** Architecture Visualized

For each method, we train and evaluate using 5-fold cross-validation. We ensure that clips from a child are maintained in one fold, as to prevent the model from artificially performing better by learning user recording idiosyncrasies (e.g., background noise). For each fold, we save the weights for the highest performing model after training and report mean accuracy (with threshold 0.5), precision, recall, F1, and AUROC, averaged over the five folds.

## Results

Out of our models, the highest-performing model was the 8-million parameter CNN model, achieving 79.3% accuracy, 80.4% precision, 79.3% recall, 79.0% F1 score, and 0.822 mean AUROC score **(Table 2)**. Our wav2vec 2.0 model performed comparably with our best CNN, achieving 76.9% accuracy, 78.2% precision, 74.6% recall, and 76.8% F1 score, and a 0.815 mean AUROC score. On the other hand, our highest performing lightweight machine learning model (Random Forest) performed somewhat worse than the other three models with 69.7% accuracy, 68.7% precision, 74.4% recall, 69.4% F1 score, and 0.740 mean AUROC score.

Our receiver operating characteristic (ROC) curves for the top three highest-performing models of each category are included in **Figure 4** subsections A, C, and E. In each figure, ROC curves for each individual fold and the mean curve are reported. One point of interest is that each figure has variation in AUC values between folds to some degree. Moreover, these variation trends are similar between models: for instance, each model appears to perform quite well on Fold 2, while performing relatively poorly on Fold 3. This suggests that the data in each fold may be too limited, resulting in folds that have differences in content that cause varying model performance from fold to fold. This disparity between AUC values is the greatest in **Figure 4** subsection A, perhaps explainable by the Random Forest classifier's small size and lightweight traits. The wav2vec model in subsection E has the most unvarying results, implying that it is better at consistently performing well at classifying unseen data than either of the other two models. This is expected, given that the wav2vec model contains far more parameters than either of the other two models and is more robust.

In **Figure 4** subsections B, D, and F, we provide confusion matrices for all three highest-performing models. D and F show that both the CNN and wav2vec models have relatively few false positive predictions, while B shows that the Random Forest classifier has a relatively large number of false positive predictions. The three figures suggest that all models have similar false negative prediction rates.

**Table 2**. Performances on *Guess What?* dataset. Results are reported with standard deviation over five different runs for each model.

|  | Accuracy | Precision | Recall | F1 | Mean AUROC |
|---|---|---|---|---|---|
| Random Forest | 0.697 ± 0.013 | 0.687 ± 0.010 | 0.744 ± 0.247 | 0.694 ± 0.013 | 0.740 ± 0.09 |
| 8M CNN | 0.793 ± 0.013 | 0.804 ± 0.014 | 0.793 ± 0.014 | 0.790 ± 0.014 | 0.822 ± 0.010 |
| wav2vec 2.0 | 0.769 ± 0.005 | 0.782 ± 0.021 | 0.746 ± 0.031 | 0.768 ± 0.006 | 0.815 ± 0.077 |

**Figure 4**. **(A)** ROC curve for Random Forest model. **(B)** Confusion matrix for Random Forest model. **(C)** ROC curve for 8M CNN. **(D)** Confusion matrix for CNN. **(E)** ROC curve for wav2vec 2.0 model. **(F)** Confusion matrix from wav2vec 2.0 model. All models were tested and trained on the *Guess What?* audio dataset, composed of child speech segments taken from educational gameplay videos.

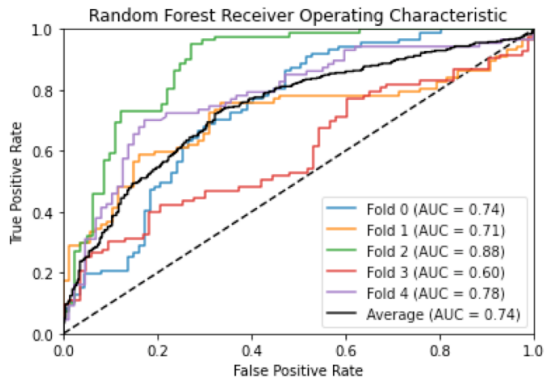

A) Random Forest ROC

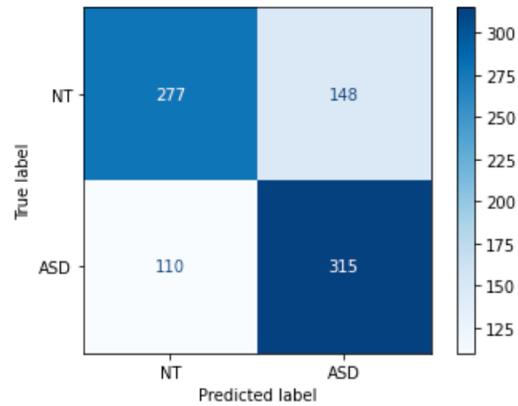

B) Random Forest Confusion Matrix

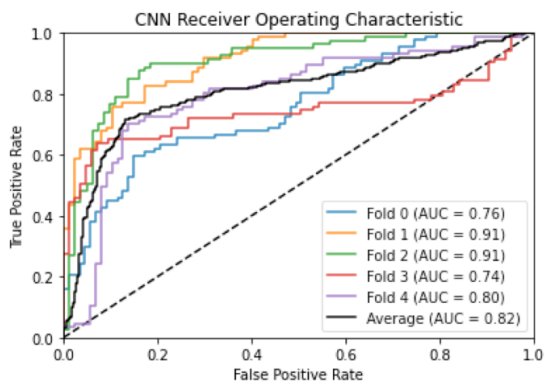

C) 8M CNN ROC

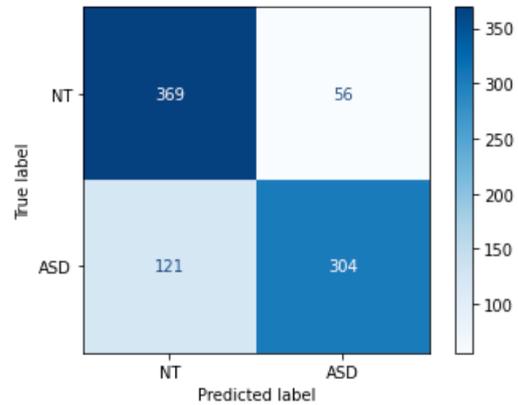

D) 8M CNN Confusion Matrix

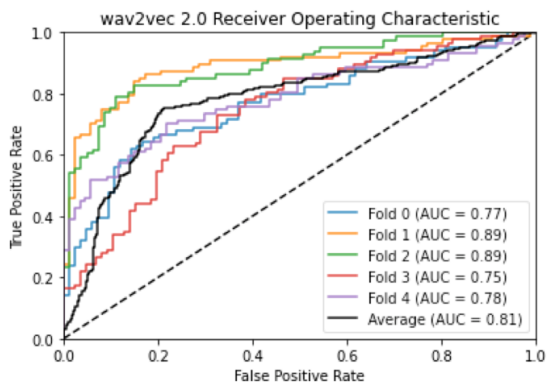

E) wav2vec 2.0 ROC

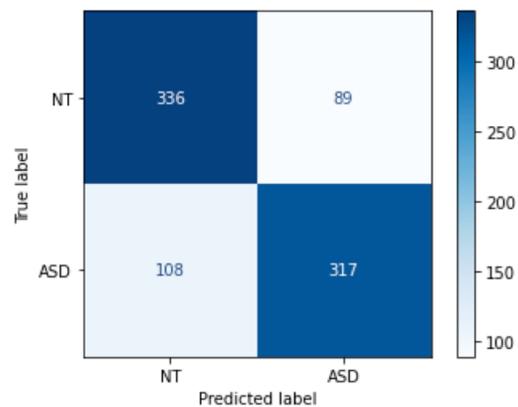

F) wav2vec 2.0 Confusion Matrix

## Discussion

### Principal Results

In this work, we trained multiple models to detect autism from our novel dataset of audio recordings curated from the educational video game *Guess What?*. We presented a set of systems which classify audio recordings by autism status and demonstrated that both CNNs and state-of-the-art speech recognition models are capable of attaining relatively high accuracies on this task, with lightweight statistical classifiers still achieving reasonable results.

### Privacy

One consideration for any recorded audio medical diagnosis is privacy [26, 28, 45], particularly important for studies involving commonly stigmatized disorders like autism [58]. We note that since our proposed models are relatively lightweight, they could be feasibly deployed at home on mobile devices, allowing for private offline screening as well as privacy-preserved federated learning approaches [27]. Prior work investigated using federated learning techniques to preserve privacy while boosting model performance on a fMRI classifier task; a similar framework might be feasible for autism diagnosis, affording a greater degree of privacy for parents who wish for a diagnosis of their child's condition but hesitate to share videos with strangers [59].

### Limitations

One limitation of our approach is the relative imbalances in the gender distribution of children who comprised our speech dataset. Our dataset included a split between 95% ASD male and 5% ASD female for autistic speech segments, as well as a 39% NT male, 58% NT female, and 3% NT unknown gender split for neurotypical speech segments. Our dataset had a sizable imbalance in terms of the relative proportion of ASD male and ASD females represented. While some imbalance is to be expected due to the naturally skewed autism sex ratio, our imbalance was larger than the observed real world 4:1 to 3:1 male-to-female incidence ratio, which would result in a dataset containing a 80-75% male and 20-25% female split for ASD segments [60, 61]. Therefore, despite being closer to replicating actual conditions than prior work, our dataset may still not be completely representative of real-world conditions.

Another limitation of our work is that we evaluate on a relatively small data set. Additionally, manually splicing videos to isolate child voices is a time-intensive process which may not be scalable to larger datasets. The alternative—automatically isolating voices through blind signal separation—is an exceptionally challenging task [62, 63]. However, it poses a potential area of interest and is possibly a necessary hurdle to overcome in order to develop widely available and consistently effective autism machine learning diagnosis resources.

## Future Work

One strength of our approach is the relatively small amount of data required to train the model. Our models were trained on clips spliced from a total of 115.5 minutes of audio yet still yielded relatively accurate results, implying that training on more data may improve performance.

Therefore, future directions include testing our models' performance with additional data from a wider selection of both autistic and neurotypical children. One particular area of interest may be wearable devices such as Google Glass [29, 30]; previous work [32, 33, 34, 35] investigated delivering actionable, unobtrusive social cues through wearables. Such approaches have been demonstrated to improve socialization among children with autism spectrum disorder [31, 36], suggesting that they could also be used to collect naturalistic data similar to this experiment in an unobtrusive way.

Another area of interest for future work may be examining the possibility of leveraging a distributed workforce of humans for extracting audio-related features to bolster screening accuracy. Previous work examined the use of crowdsourced annotations for autism, indicating that similar approaches could perhaps be applied through audio [39-45]. Combining the audio feature extraction with other autism classifiers could be used to create an explainable diagnostic system [37, 38, 46-57] fit for mobile devices [52]. Previous work investigated using such classifiers to screen for autism or approach autism-related tasks like identifying emotion to improve socialization skills; combining computer vision-based quantification of relevant areas of interest, including hand stimming [50], upper-limb movement [56], and eye contact [55, 57], could possibly result in interpretable models.

## Conclusions

Screening for autism with an automatic audio classifier could possibly accelerate the lengthy diagnosis process. Our models were able to predict autism status by training on a varied selection of home audio clips with inconsistent recording qualities and are more generalizable to real world conditions. Overall, our work suggests a promising future for at-home screening for autism spectrum disorder.

## Acknowledgements

This work was supported in part by funds to DPW from the National Institutes of Health (1R01EB025025-01, 1R21HD091500-01, 1R01LM013083, 1R01LM013364), the National Science Foundation (Award 2014232), The Hartwell Foundation, Bill and Melinda Gates Foundation, Coulter Foundation, Lucile Packard Foundation, the Weston Havens Foundation, and program grants from Stanford's Human Centered Artificial Intelligence Program, Stanford's Precision Health and Integrated Diagnostics Center (PHIND), Stanford's Beckman Center, Stanford's Bio-X Center, Predictives and Diagnostics Accelerator (SPADA) Spectrum, Stanford's Spark Program in Translational Research, Stanford mediaX, and Stanford's Wu Tsai Neurosciences Institute's Neuroscience: Translate Program. We also acknowledge generous support from David Orr, Imma Calvo, Bobby Dekesyer and Peter Sullivan. P.W. would like to acknowledge support from Mr. Schroeder and the Stanford Interdisciplinary Graduate Fellowship (SIGF) as the Schroeder Family Goldman Sachs Graduate Fellow.

## Conflicts of Interest

D.P.W. is the founder of Cognoa.com. This company is developing digital health solutions for pediatric healthcare. All other authors declare no competing interests.

## Abbreviations

ASD: autism spectrum disorder
NT: neurotypical